# Enhancing Understanding of Hydraulic Fracture Tip Advancement through Inversion of Low-Frequency Distributed Acoustic Sensing Data

**Yongzan Liu[1], Lin Liang[1], and Smaine Zeroug[1]**

[1]Schlumberger-Doll Research, Cambridge, Massachusetts, USA.

Corresponding author: Yongzan Liu (yliu133@slb.com)

**Key Points:**

- Low-frequency distributed acoustic sensing provides a unique dataset to characterize the fracture propagation process.
- Details about the propagation dynamics and geometry of fluid-driven fractures can be inferred from distributed acoustic sensing data.
- A gradient-based inversion algorithm is developed and validated using a synthetic case to estimate the fracture tip advancing process.
- In the presented field case, fracture propagates continuously in the beginning, followed by an intermittent tip advancement pattern, which is a unique field evidence of fracture tip behaviors.

**Abstract**

Characterizing the fluid-driven fracture tip advancing process presents a significant challenge due to the difficulty of replicating real-world conditions in laboratory experiments and the lack of precise field measurements. However, recent advances in low-frequency distributed acoustic sensing (LF-DAS) technology offer new opportunities to investigate the dynamics of propagating hydraulic fractures. In this study, we propose an iterative inversion method to characterize fracture-tip advancing behaviors using LF-DAS data. A forward geomechanical model is developed using the three-dimensional displacement discontinuity method, and the optimization is realized by a conjugate gradient method. The performance of the inversion algorithm is demonstrated using a synthetic case, in which the fracture half-length evolution and propagation velocity match well with the reference solutions. Additionally, the averaged fracture cross-section area, fracture volume, and fracturing fluid efficiency can also be estimated, showing good agreements with true values of the synthetic case under reasonable assumptions. Then a field case with a single-cluster hydraulic fracturing treatment from the Hydraulic Fracturing Test Site 2 project (HFTS2) is studied. Our analysis of the inversion results reveal that the fracture propagates intermittently, as evidenced by the fracture half-length evolution. This unique field evidence can guide modeling efforts to incorporate this important physical behavior into fracture models, and the secondary information gathered from the study, including fracture cross-section area and volume, can help evaluate and optimize fracturing efficiency.

**1 Introduction**

Hydraulic fracturing has become the primary method for developing tight/shale hydrocarbon reservoirs and is also critical to the production of clean energy. For example, this technique is used to recover heat energy through cold water injection in deep hot granite, where fractures need to be created to enable fluid flow (McLennan et al., 2023). However, the fluid-driven fracture propagation in deep geological reservoirs is a complex phenomenon, involving multi-scale multi-physical processes such as fluid flow, rock deformation, crack initiation and evolution, and even heat transport in non-isothermal conditions. Directly observing the dynamic process in the subsurface is difficult, which presents a major challenge to fully understand the fracture propagation in geological formations. Laboratory experiments and numerical modeling are two commonly used approaches to investigate the influencing factors on fracture growth. Lhomme et al. (2002) conducted hydraulic fracturing experiments with acoustic monitoring in sandstone under various fluid viscosities and injection rates. Intermittent fracture advancement was observed under scenarios with low fluid viscosity and high injection rate. Similar observations were also observed in Mode-I crack propagation in hydrogels (Pizzocola et al., 2013) and acrylic blocks (Tinni et al., 2019). A recent micro-scale experimental study by Cochard et al. (2023) reported unexpected fracture front propagation dynamics: a stick-break instability that means the propagation of the fracture front is discontinuous with no-motion pauses followed by rapid forward advancement. However, not all laboratory-scale experiments in literature exhibited obvious stepwise fracture tip advancement (Medlin and Masse, 1984; Bunger and Detournay, 2008; Chen et al., 2015; Lecampion et al., 2016; Liu and Lecampion, 2022). Laboratory experiments usually have limited scales that cannot fully replicate what happens in the field, which could be one reason for different observations. Numerical simulation is a better option to set up more realistic models, but the predictions are influenced by the assumptions adopted in different methods, especially on the 'tip element' that controls the propagation (Adachi et al., 2007). Analytical solutions, such as well-known PKN and KGD models (Perkins and Kern, 1961, Geertsma and De Klerk, 1969), and tip-asymptotic-based models (Garagash and Detournay, 2000, Peirce and Detournay, 2008) yield smooth fracture front evolution. Continuous fracture tip advancement is also obtained by most numerical solutions, mainly due to the adopted fracture footprint tracking strategies, time-stepping schemes, mesh resolutions, etc. (Peruzzo et al., 2019). Lecampion et al. (2018) and Chen et al. (2022) conducted comprehensive reviews on the advances and limitations of different modeling approaches. On the contrary, a few simulation studies reported the intermittent or stepwise fracture advancing phenomenon using different numerical methods (Schrefler et al., 2006, Secchi and Schrefler, 2014, Milanese et al., 2016, Cao et al., 2018). Peruzzo et al. (2019) stressed that stepwise fracture tip advancement does exist and has been proved with laboratory experiments and indirect field observations (e.g., wellbore pressure), which cannot be ignored in numerical models. However, previously, it is unfortunately difficult to obtain direct field observations or measurements about the evolution of fractures in the subsurface geological formations (Secchi and Schrefler, 2014).

Recently, a new fiber-optic-based technology, named low-frequency distributed acoustic sensing (LF-DAS), has been successfully deployed in the field to monitor the hydromechanical responses in the subsurface. DAS utilizes Rayleigh scattering to measure the strain-rate responses along the entire fiber, which can provide continuous high-resolution data in real-time. An interrogator is connected to one end of a fiber-optic cable on the surface and injects laser pulses successively into the fiber. The backscattered signals are processed with interferometric techniques to obtain dynamic strain rates over a specific gauge length that represents the spatial resolution. Hartog (2017) and Lindsey et al. (2020) present details on the sensing principles of DAS. The low-frequency components of DAS data are sensitive to temperature and strain variations. Becker et al. (2017) monitored strain responses using ultra-low frequency distributed acoustic sensing during a periodic hydraulic test. Jin and Roy (2017) were the first to discuss the potential of cross-well LF-DAS data for hydraulic fracture monitoring and characterization in the oil and gas industry. Recent studies have demonstrated the success of the LF-DAS technology in hydraulic fracture monitoring and diagnostics. Many high-quality field data were acquired from different formations (Ichikawa et al., 2019, Li et al., 2020, Liu et al., 2022, Ugueto et al., 2019; 2021). Several forward modeling studies were conducted by different researchers to help interpret the observed LF-DAS signals (Sherman et al., 2019; Liu et al., 2020; Zhang et al., 2020; Tan et al., 2021; Wang et al. 2022; Srinivasan et al., 2023a). A common conclusion is that the LF-DAS data is directly related to the fracture propagation process and resulting fracture geometries. In addition, Liu et al. (2021a; 2021b) developed a first-of-its-kind inversion algorithm to estimate the fracture width and height in the vicinity of the monitoring well after fracture hit. The robustness and accuracy of this algorithm have been tested with both synthetic cases (Liu et al. 2021a; 2021b) and field cases with hundreds of fracturing stages (Liu et al. 2022; Srinivasan et al. 2023b). However, the full potential of this novel data remains untapped, as existing studies have largely overlooked the LF-DAS data before fracture hit. We hypothesize that the portion of data before fracture hit captures the evolution of fracture tip, which can provide field evidence on how the fracture propagates in the subsurface.

In this study, we introduce a new approach to estimating fracture propagation characteristics using low-frequency distributed acoustic sensing. Specifically, we propose a gradient-based inversion algorithm that utilizes direct far-field fracture-induced strain variations to estimate the evolution of the fracture tip. Secondary fracture parameters, such as averaged fracture cross-section area, fracture volume and fracturing fluid efficiency, can be estimated under certain assumptions. Our method represents a novel approach to analyzing LF-DAS data before fracture hit and has the potential to provide unique insights obtained directly from precise field measurements into the mechanics of hydraulic fracturing. In the following sections, we firstly present the detailed methodology, followed by the validation and performance evaluation with a synthetic case. Lastly, we present a field case study illustrating how fracture tip advances under certain conditions.

## 2 Methodology

To harness the significant and unique far-field strain responses induced by propagating fractures, we propose a gradient-based inversion method for characterizing the advancement of fracture tips. In order to develop the forward model, we have implemented the displacement discontinuity method (Crouch, 1976; Shou, 1993), a well-established modeling approach in the field of hydraulic fracturing. This method enables us to relate the fracture geometry to the strain measurements obtained from low-frequency distributed acoustic sensing along a horizontal wellbore. As depicted in Figure 1, the monitoring well can continuously record the evolving strain profiles induced by the propagating fracture. In the forward model, we incorporate the ability to account for the varying fracture width across different locations. To achieve this, we discretize the fracture into sub-elements, each characterized by a height of $2h$, a length of $2l_f$, and a width of $w$. This allows us to represent the geometry of the fracture more realistically at different positions along its propagation path.

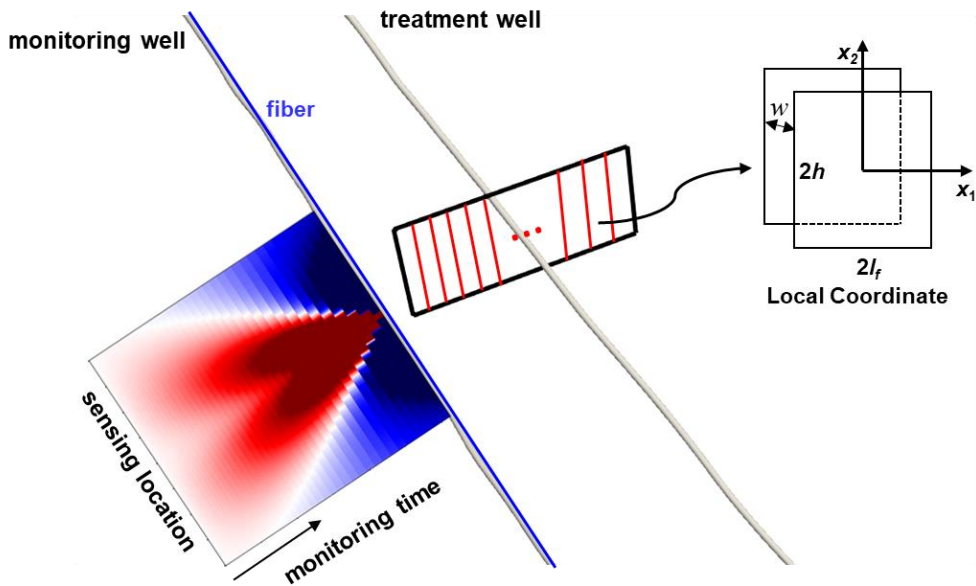

Figure 1. Schematic of a well pair consisting of a treatment well and a monitoring well. The waterfall (contour) plot shows the fracture-induced strain evolution that can be measured by LF-DAS. The fracture can be discretized into small elements to accommodate the varying fracture width along the fracture.

At each timestep, considering tensile fractures, the directional strain ($\boldsymbol{\varepsilon}$) along the wellbore can be calculated as:

$$\boldsymbol{\varepsilon} = \mathbf{G}\mathbf{w} \tag{1}$$

where **G** represents the Green-function matrix that is dependent on the fracture height ($H$), fracture length ($L$), and the relative positions of the fracture and sensing points. **w** is the fracture width vector. For a fracture with $N$ elements, the fracture length $L$ can be expressed as $L = \sum_{j=1}^{N} 2l_f$. The fracture width distribution (**w**) can be further constrained based on some physical understanding of the fracture geometry, e.g., constant, elliptical, penny-shaped distribution, etc. Therefore, the width distribution vector (**w**) can be expressed as a function of width at the perforation point ($w_0$) and a shape operator (**S**), i.e., $\mathbf{w} = \mathbf{S}w_0$. Equation (1) can be written as:

$$\boldsymbol{\varepsilon} = \mathbf{GS}w_0. \tag{2}$$

The key component of the forward model is to calculate the **G** matrix. DDM is a fast and accurate method for fracture problems in elastic bodies (Crouch, 1976; Shou, 1993). In a local coordinate as shown in Figure 1, the displacements at any arbitrary location $\mathbf{x} = (x_1, x_2, x_3)$ induced by a fracture element $j$ with displacement discontinuities $D_i$, $i = 1, 2, 3$ are expressed as:

$$\begin{aligned}
u_1^j &= 2(1-v)\Phi_{1,3} - (1-2v)\Phi_{3,1} - x_3\Phi_{k,k1} \\
u_2^j &= 2(1-v)\Phi_{2,3} - (1-2v)\Phi_{3,2} - x_3\Phi_{k,k2} \\
u_3^j &= 2(1-v)\Phi_{3,3} - (1-2v)(\Phi_{1,1} + \Phi_{2,2}) - x_3\Phi_{k,k3}
\end{aligned} \tag{3}$$

where $v$ is the Poisson's ratio. The notation ',' represents the partial derivative. The functions $\Phi_i$ are related to the displacement discontinuities, which is written as:

$$\Phi_i = \frac{1}{8\pi(1-v)} D_i I(x_1, x_2, x_3), \tag{4}$$

where

$$I(x_1, x_2, x_3) = \int_{-h}^{h} \int_{-l_f}^{l_f} r^{-1} d\xi_1 d\xi_2, \tag{5}$$

and

$$r = \sqrt{(x_1 - \xi_1)^2 + (x_2 - \xi_2)^2 + x_3^2}. \tag{6}$$

The explicit form of equations for $I$ and its derivatives can be found in Liu (2021). After obtaining the displacements in the local coordinate, we can transform them into the global coordinate by:

$$\begin{bmatrix} u_x^j \\ u_y^j \\ u_z^j \end{bmatrix} = \begin{bmatrix} \cos\beta & -\sin\beta & 0 \\ 0 & 0 & 1 \\ \sin\beta & \cos\beta & 0 \end{bmatrix} \begin{bmatrix} u_1^j \\ u_2^j \\ u_3^j \end{bmatrix}, \tag{7}$$

where $\beta$ is the angle between the global coordinate and local coordinate. Then the strain induced by the whole fracture can be obtained by superimposing the contributions from all the boundary fracture elements:

$$u_x = \sum_{j=1}^{N} u_x^j, \quad u_y = \sum_{j=1}^{N} u_y^j, \quad u_z = \sum_{j=1}^{N} u_z^j. \tag{8}$$

Finally, the strain at sensing point $k$ along the monitoring well, i.e., components of $\boldsymbol{\varepsilon}$ in Equation (1), can be calculated following:

$$\varepsilon_k = \frac{u_{k+L/2} - u_{k-L/2}}{L}, \tag{9}$$

where $L$ represents the gauge length. $u_k$ denotes the displacement projected along the monitoring well direction at sensing location $k$. $\varepsilon_k$ is the averaged strain over $L$, which is equivalent to the LF-DAS measurement resolution (Liu et al., 2020; 2021c).

In the proposed method, we assume a constant fracture height ($H$). Therefore, the problem is simplified as the optimization of $w_0$ and $L$ to minimize the residual between modeled strain ($\boldsymbol{\varepsilon}$) and field strain ($\mathbf{d}$) within a bounded region defined by $w_{0,\min}$, $w_{0,\max}$, $L_{\min}$ and $L_{\max}$, written as:

$$\min\{f(L, w_0): L_{\min} \leq L \leq L_{\max}, w_{0,\min} \leq w_0 \leq w_{0,\max}\}, \tag{10}$$

where $f = \mathbf{G}\mathbf{S}w_0 - \mathbf{d}$. Because $\mathbf{G}$ depends on the unknown parameter $L$, it is nontrivial to explicitly calculate $\mathbf{G}$. Iterative optimization with an initial guess on the model parameters is recommended. In this study, a subspace, interior, and conjugate gradient method developed by Branch et al. (1999) is applied to solve the optimization problem. Based on the inversion results, i.e., $L$ and $w_0$, we further calculate the fracture propagation velocity and fracture surface/volume.

## 3. Inversion Algorithm Performance Evaluation

In this section, we assess the effectiveness of our inversion algorithm through the examination of a synthetic case, which is adapted from Liu et al. (2021a). Figure 2(a) presents the profile of the fracture half-length throughout the fluid injection period. The propagation velocity is determined by dividing the increment in fracture length between two consecutive time steps by the corresponding time interval. Moreover, the fracture has a height of 21.3 m. Figure 2(b) illustrates the synthetic LF-DAS strain data, which is recorded along an offset well located at a distance of 100 m from the treatment well. In this study, our focus lies on the data preceding the fracture hit, which is delineated by the dashed lines. The characteristic pattern is a 'sandwich' configuration, wherein a heart-shaped extensional zone (red color) is enclosed by compressional zones (blue color). Following the fracture hit, the strain data exhibits higher sensitivity to the opening and closing of the fracture in proximity to the monitoring well. The primary distinction in the signal pattern is the contraction of the heart-shaped extensional zone, which transforms into a broader band encircling the fracture-hit location. The width of this red band is dependent on the gauge length, which is about 10 m in this synthetic case. For a more comprehensive analysis and interpretation of the qualitative LF-DAS signatures, we refer readers to the works of Liu et al. (2020; 2021c). Further details regarding the synthetic case can be found in Liu et al. (2021a). The primary objective of this synthetic case study is to conduct a comparison of the inverted fracture tip evolution and propagation velocity against the reference solutions shown in Figure 2(a). By doing so, we aim to evaluate the accuracy and reliability of our inversion approach. Furthermore, we showcase the potential of the inversion approach in estimating additional fracture parameters, including fracture cross-section area, fracture volume, as well as fracturing efficiency. These parameters play a crucial role in characterizing the behavior and effectiveness of the fracture propagation process.

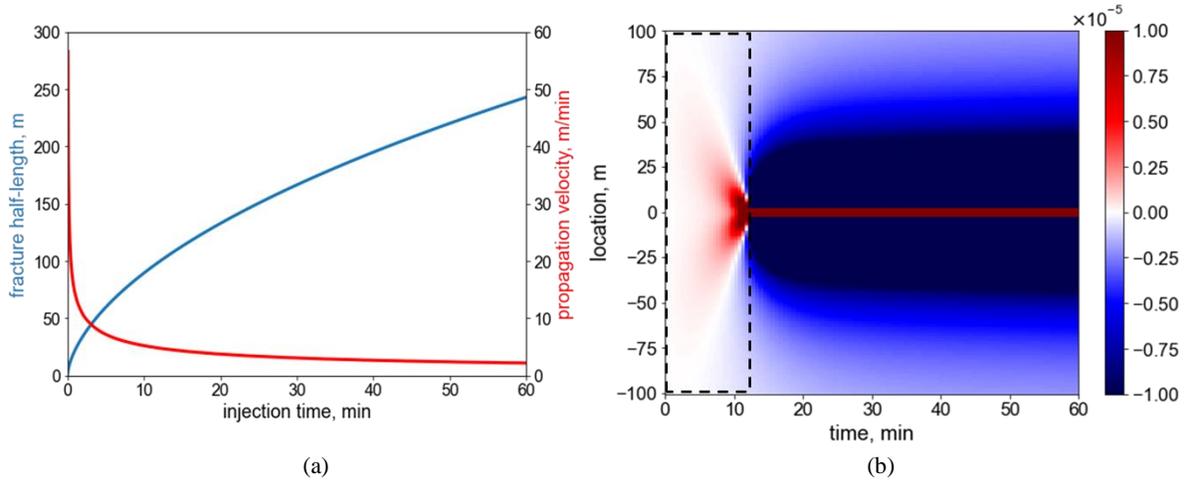

(a)　　　　　　　　　　　　　　　　　　(b)
Figure 2. (a) fracture half-length and propagation velocity as a function of injection time; (b) waterfall plot of synthetic LF-DAS strain data along a monitoring well with 100 m offset distance to the treatment well.

We set the element size to be 2 m. The lower bounds for fracture half-length and fracture width are 0.5 m and 0.001 mm, respectively, while the upper bounds are 150 m and 10 mm. These bounds help define a realistic range within which the inversion algorithm can operate effectively and provide meaningful results. In Figure 3(a), we present a comparison between the inverted fracture half-length and propagation velocity against their true values. Additionally, Figure 3(b) compares the modeled strain data and the synthetic data. The excellent agreements observed in these comparisons indicate that the modeled strain converges to the input strain data, while the inversion results demonstrate minimal discrepancies when compared to the true solutions. Additionally, based on the inversion results, specifically the fracture length ($L$) and width distribution ($\mathbf{w}$), we can calculate the averaged fracture cross-section area ($\bar{w} \times H$) and fracture volume ($\bar{w} \times H \times L$). Here, $\bar{w}$ represents the average fracture width, and $H$ denotes the fracture height. These calculations provide us with valuable insights into the overall size and volume of the fractured zone, allowing for a more comprehensive understanding of the fracturing process and its efficiency. Figure 4(a) compares the calculated fracture cross-section area against the true solution. It demonstrates good agreements during the early injection period. However, some deviations become apparent at later times. These discrepancies arise from the assumption made regarding the fracture width distribution. As shown in Figure 5(a), the inverted width distributions exhibit a strong resemblance to the true width profiles during the initial times. However, notable differences emerge at later stages. For instance, at 11.863 minutes (represented by the purple color), only the widths near the fracture tip exhibit a satisfactory match with the true widths. Nevertheless, it is worth noting that the strain profiles consistently exhibit a high level of agreement with the true data, as illustrated in Figure 5(b). The observed phenomenon can be attributed to the fact that as the fracture approaches the monitoring well, the strain is predominantly influenced by the fracture segments near the tip. These segments induce extensional strain, which is one order higher than the surrounding compression, as demonstrated by the purple plots in Figure 5(b). Although the mismatch in the fracture width distribution does not significantly affect the prediction of fracture advancement, it can result in inaccurate estimations of fracture area/volume. To address this issue, reducing the search range of fracture width during the inversion process may help alleviate the discrepancies and improve the accuracy of the estimated fracture parameters. By narrowing down the potential range of fracture widths, the inversion algorithm can focus on capturing the width distribution over the whole fracture. Figure 6 shows the width profiles and the averaged fracture area when a reduced upper bound for the fracture width is employed. This adjustment leads to improved inversion results at the later times by constraining the fracture width closer to the true value. The analysis indicates that the assumption made on the fracture width distribution does not significantly impact the estimations of fracture length and propagation velocity. However, to obtain a reasonable estimate of these secondary fracture parameters, it may be necessary to conduct multiple inversions with different upper bounds for the fracture width. This approach allows for a more comprehensive exploration of the potential width distributions and improves the accuracy of these secondary parameters.

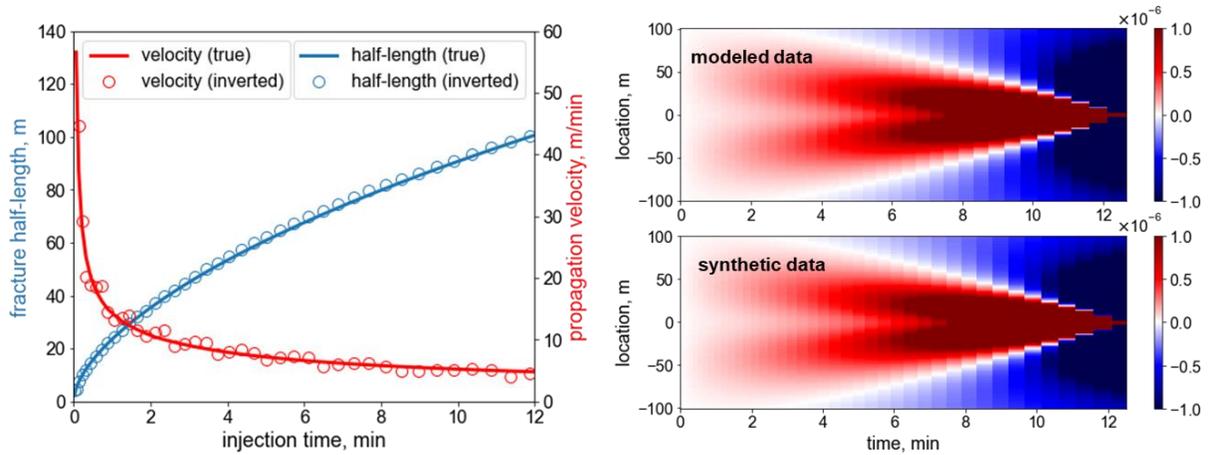

Figure 3. (a) comparison of fracture half-length and propagation velocity between inversion results and reference solutions; (b) comparison between modeled strain data and input synthetic data.

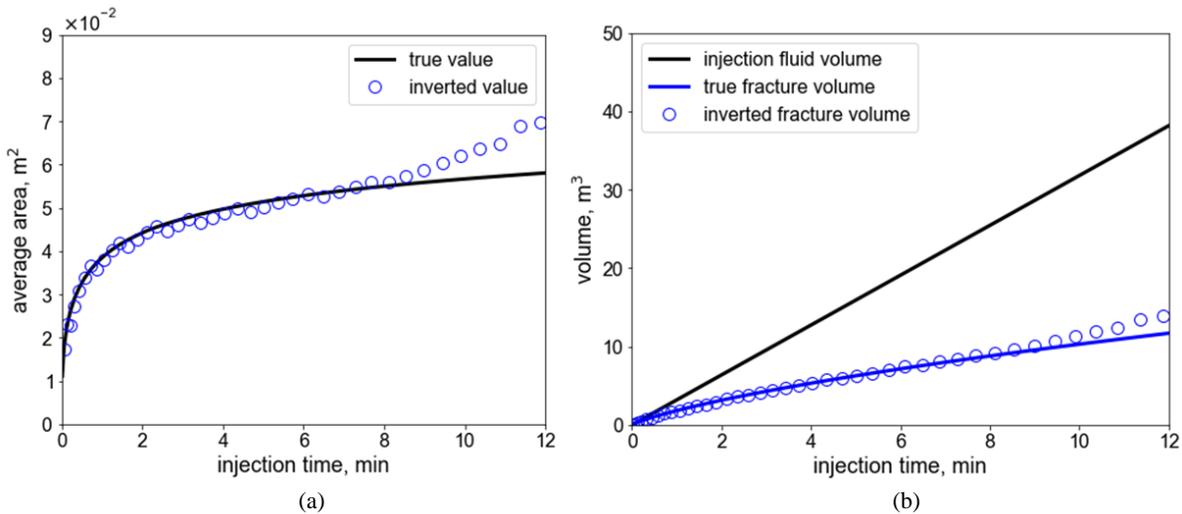

(a) (b)

Figure 4. Comparisons of (a) averaged fracture cross-section area and (b) fracture volume together with fluid volume as a function of injection time.

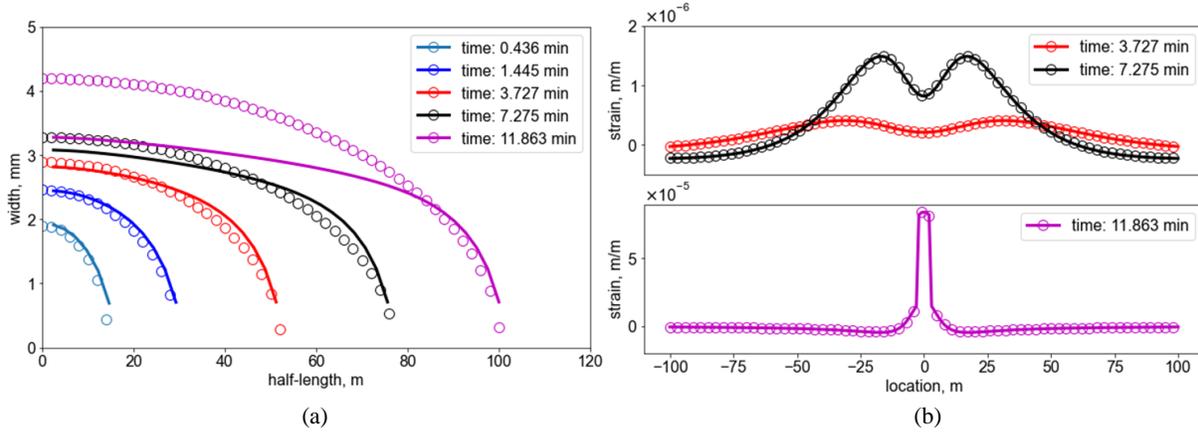

(a)          (b)

Figure 5. (a) fracture width profiles at different times; (b) comparisons of modeled strain and true strain at three specific times. Dots represent inversion results and solid curves represent reference solutions.

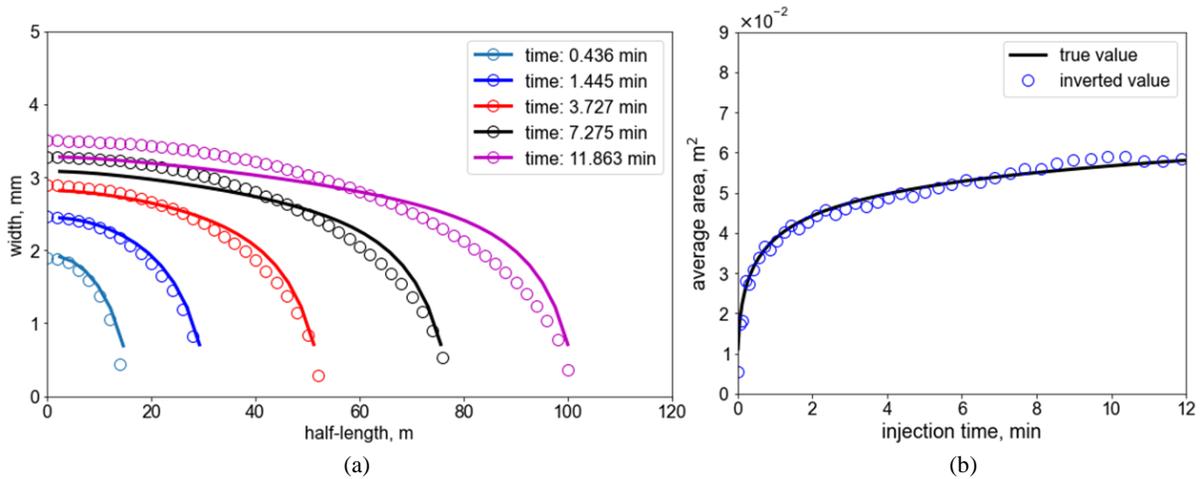

(a)          (b)

Figure 6. Comparisons of (a) fracture width profiles at different times; (b) averaged fracture cross-section area. Results are obtained by reducing the upper bound of the fracture width to 0.35 mm. Dots represent inversion results and solid curves represent reference solutions.

In the proposed method, the fracture height is assumed to be a constant value. However, in practice, the precise fracture height may be not readily available. Therefore, it is essential to investigate the sensitivity of the inversion results to the fracture height. We conducted additional inversions using two different heights: twice of the true height (42.6 m) in one case and half of the true height (10.65 m) in the other case. Figure 7 compares the evolutions of estimated fracture half-length and propagation velocity obtained with these deviated heights against the true solutions. Overall, the inversion results from both cases demonstrate good agreement with the true solutions. This suggests that the pre-assumed fracture height does not significantly impact the estimation of fracture tip advancement. The preceding analysis indicates that the proposed method is robust and reliable, as it can accurately predict fracture propagation regardless of minor variations in the assumed fracture height.

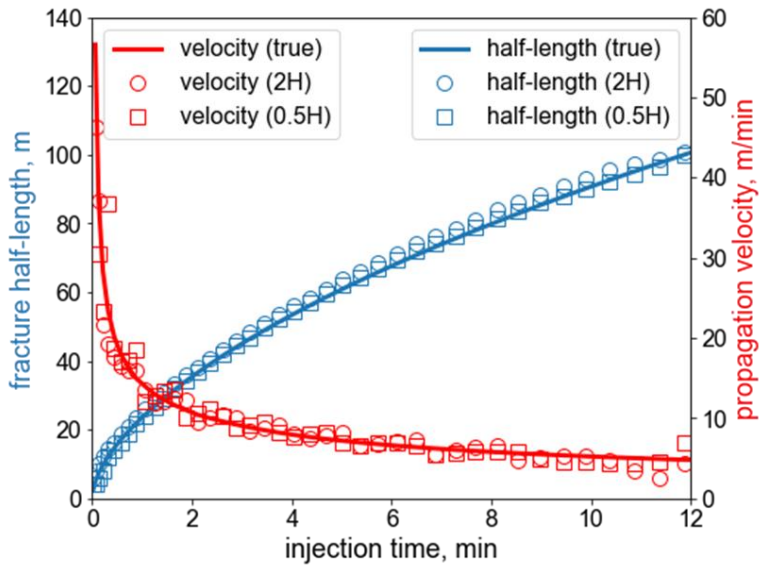

Figure 7. Comparisons of fracture half-length and propagation velocity as a function of injection time between the reference solutions and inversion results under two different heights.

## 4 Field Case Study

The Hydraulic Fracturing Test Site-2 (HFTS-2) project is a field-scale research and development program that focuses on multidisciplinary field data acquisition and interpretation to gain better understanding of hydraulic fracture propagation and resulting fracture geometry (Zhao et al., 2019). The broad deployment of fiber-optic cables in HFTS-2 is a unique characteristic compared to previous scientific pilot experiments. A few simple single-cluster stages were designed with the hope to gather direct information on the propagating fracture by eliminating the effects of fracture interaction in spontaneous multi-fracture propagation. In this section, we applied the proposed inversion approach to one stage with a single-cluster design from HFTS-2. In the following, we first briefly describe the field measurements and then focus on the analysis of the inversion results.

4.1 Field Measurement Description

As shown in **Figure 8**, the field LF-DAS strain-rate data used in this study is monitored along horizontal well B3H during a hydraulic fracturing job in well B4H. B5H is a vertical monitoring well and the data from B5H contains information about fracture height, which will be discussed in the next section. The well spacing between the two wells is about 200 m and B4H is about 64 m above B3H in the vertical direction. The fracture hit is identified at 9:04, 18/Apr/2019 following the guideline described in Section 3. Before the fracture hit (9:04, 18/Apr/2019), the LF-DAS data represents the strain responses in front of the advancing fracture tip.

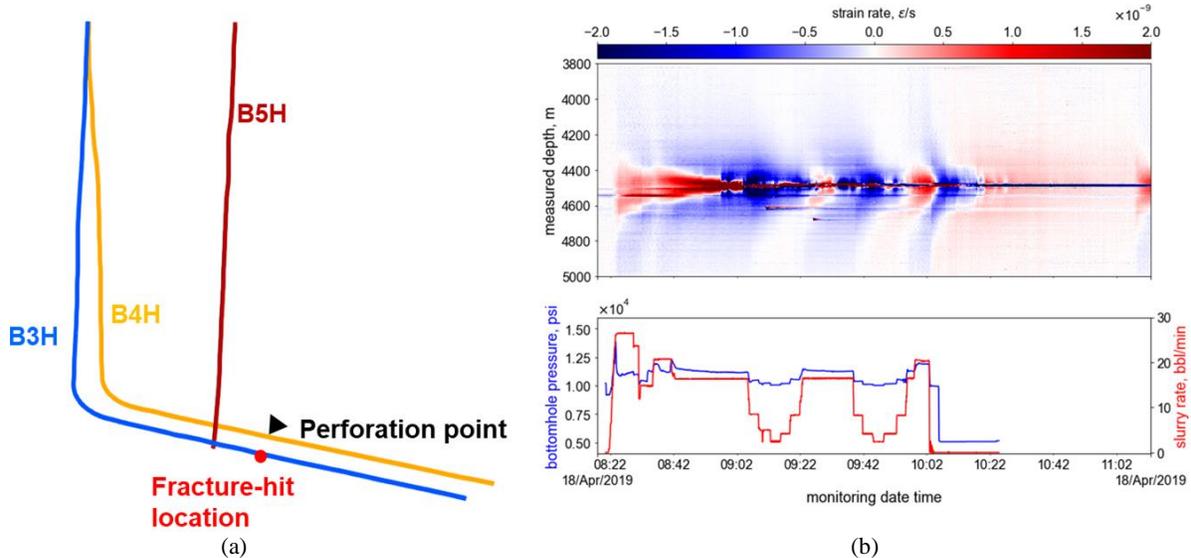

(a)                           (b)

**Figure 8.** Configuration of a well pair consisting of a treatment well and a monitoring well (a) and LF-DAS strain-rate waterfall plot together with the pumping curves (b). In (a), B4H is the treatment well and B3H is the monitoring well installed with fiber. The black triangle and red dot indicate the perforation point and fracture-hit location respectively. The top waterfall plot in (b) illustrates the LF-DAS strain-rate data monitored along a horizontal section of B3H. The bottom plot in (b) shows the bottomhole pressure and slurry rate of the stimulation job.

Our focus in this paper is on estimating fracture tip advancement through inversion of LF-DAS strain data prior to fracture hit. **Figure 9**(a) depicts a smaller time window with strain and strain-rate data, where the strain data is obtained by integrating the field strain-rate data over time. This serves as the 'data', i.e., **d** in Equation (10), for the inversion algorithm. The vertical black line indicates the fracture-hit time. Intuitively, the accuracy of the inversion results can be verified by comparing the fracture half-length on the side of the monitoring well to the well spacing at the fracture-hit time. In **Figure 9**(b), we present the LF-DAS measurements from a nearby vertical well, specifically B5H in **Figure 8**(a), taken during the early pumping period prior to the fracture reaching the horizontal monitoring well B3H.

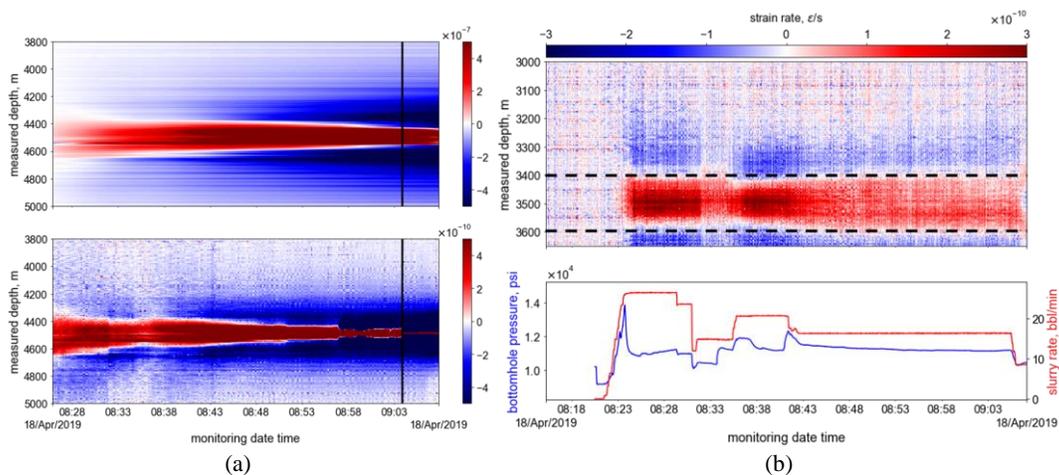

(a)                           (b)

**Figure 9.** (a) Waterfall plots of LF-DAS strain (top plot) and strain rate (bottom plot) before 09:08, 18/Apr/2019. The strain data is obtained by integrating the strain rate data over time. (b) LF-DAS strain rate responses from a nearby vertical monitoring well (B5H in **Figure 8**(a)) during the early stage of a hydraulic fracturing job. The two black dash lines label the boundaries of the red extending zone.

4.2 Inversion Result Analysis

We initiate the inversion process at 08:26, 18/Apr/2019, about 5 minutes after the commencement of fluid injection. This was done to take advantage of the stronger and clearer LF-DAS measurements obtained from the offset monitoring well at this time. The inversion is terminated at 09:03, 18/Apr/2019, just prior to the fracture reaching the monitoring well. The units for fracture width at the perforation point and fracture half-length are in meters, with bounds of $[1.0\times10^{-6}, 1.0\times10^{-2}]$ and $[0.5, 250]$, respectively. At the first timestep, we use initial guesses of $2.0\times10^{-4}$ m and 20 m for the width and half-length, respectively. The initial guess for each subsequent time step is set as the inversion result from the previous time step. As described in the Methodology section, the proposed inversion algorithm requires an assumed fracture height, which can be inferred from LF-DAS data monitored along vertical wells (Zhang et al., 2020). The boundaries of the fracture height can be determined by the transitions from the extending (red) to compressing (blue) labeled by the black dashed lines in **Figure 9**(b), according to the comprehensive forward modeling study by Zhang et al. (2020) and Srinivasan et al. (2023a). Based on this, it can be inferred that the fracture height during the early period shown in **Figure 9**(b) is approximately 200 m.

Although the fracture height is inferred from reliable field measurements, it is still an approximate value that may deviate slightly from the true value. Although the sensitivity has been tested using a synthetic case, it is still necessary to examine the impact of pre-defined fracture height on the inversion performance of field cases with noisy measurements. Therefore, we construct 3 cases with different fracture heights, i.e., 150 m, 200 m, and 250 m. The results of sensitivity analysis are presented in **Figure 10**, which shows the temporal evolutions of fracture half-length and propagation velocity for the three cases with different heights. The lack of noticeable differences in the inversion results among the three cases indicates that certain deviations in the fracture height do not significantly impact the inversion performance in field applications. **Figure 11**(a) qualitatively compares the field strain data with the modeled strain data under different fracture heights for the entire monitoring period from the start of inversion to just before the fracture reaching the monitoring well. **Figure 11**(b) quantitatively compares the spatial strain profiles at four specific times. Clearly, all the modeled strain data from the three cases match very well with the field data, which demonstrates the validity of the inversion results and the insensitivity to fracture height.

Next, we focus on analyzing the fracture-tip advancement characteristics based on the inversion results. In **Figure 10**, the fracture propagation velocity is calculated by dividing the length difference between two adjacent time steps by the time interval. In this study, the timestep size is 30 seconds. Before 08:30, 18/Apr/2019, the fracture length generally grows smoothly with the propagation velocity decreasing from over 20 m/min to about 5 m/min. Between 08:30 and 08:41, the velocity shows a more pronounced drop followed by a slight increase, which is corresponding to the injection rate/pressure variations, as shown in **Figure 9**(b). After 08:41, the injection rate is almost constant, and the bottom-hole pressure also remains stable. Interestingly, the fracture advances in an intermittent/stepwise manner, as clearly illustrated in **Figure 10**. In this specific case, there are four cycles of stepwise advancement after 08:41, in which the fracture advances for a certain period followed by a quiescent period with almost zero velocity. It is interesting to note that the duration of the quiescent period tends to increase as the fracture propagates further. This observation can be explained by the coupled hydromechanical nature of fluid-driven fracture propagation in saturated porous media. When a new fracture surface is created, the increased fracture volume can induce pressure reduction near the fracture tip. As the fracturing fluid continues to be pumped into the formation, the fluid pressure in the near-tip region will increase. When the pressure reaches a critical value, a new fracture surface is created followed by another quiescent period. The duration of the quiescent period depends on the rate at which the injected fluid can increase the fracture pressure in the near-tip region. The pressure transient process along the fracture is influenced by various factors, including fracture length, matrix permeability (leak-off), fracture conductivity, fluid properties, and operational parameters such as injection rate. These factors affect the fluid flow in the fracture and the surrounding matrix, leading to pressure changes that can impact the fracture-tip advancement behavior. For instance, when the injection rate is constant, an increase in fracture length leads to more fluid leak-off into the surrounding formation and a longer pressure travel time. Therefore, it takes longer for the pressure to build up to the fracturing condition, which could explain why the quiescent period increases as the fracture propagates. The direct field evidence of such behavior from LF-DAS measurements not only confirms the fracture-tip advancing pattern but also provides opportunities to improve theoretical models by incorporating essential physical processes in the near-tip region and to develop schemes to optimize the fracturing job itself in any specific formation being monitored with LF-DAS.

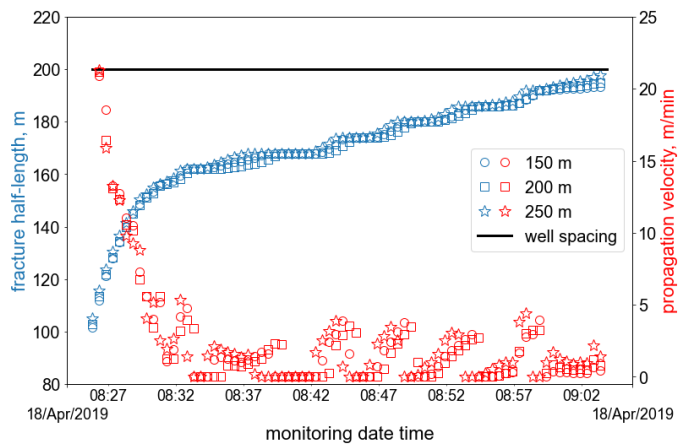

**Figure 10.** Fracture half-length (left y-axis) and propagation velocity (right y-axis) as a function of monitoring time. This latter features an intermittent behavior dropping to zero and rising again as the fracture propagates towards the monitoring well. The black line indicates the well spacing. Right before the fracture-hit time, the inverted fracture half-length is very close to the well spacing which indicates the inversion results are reliable.

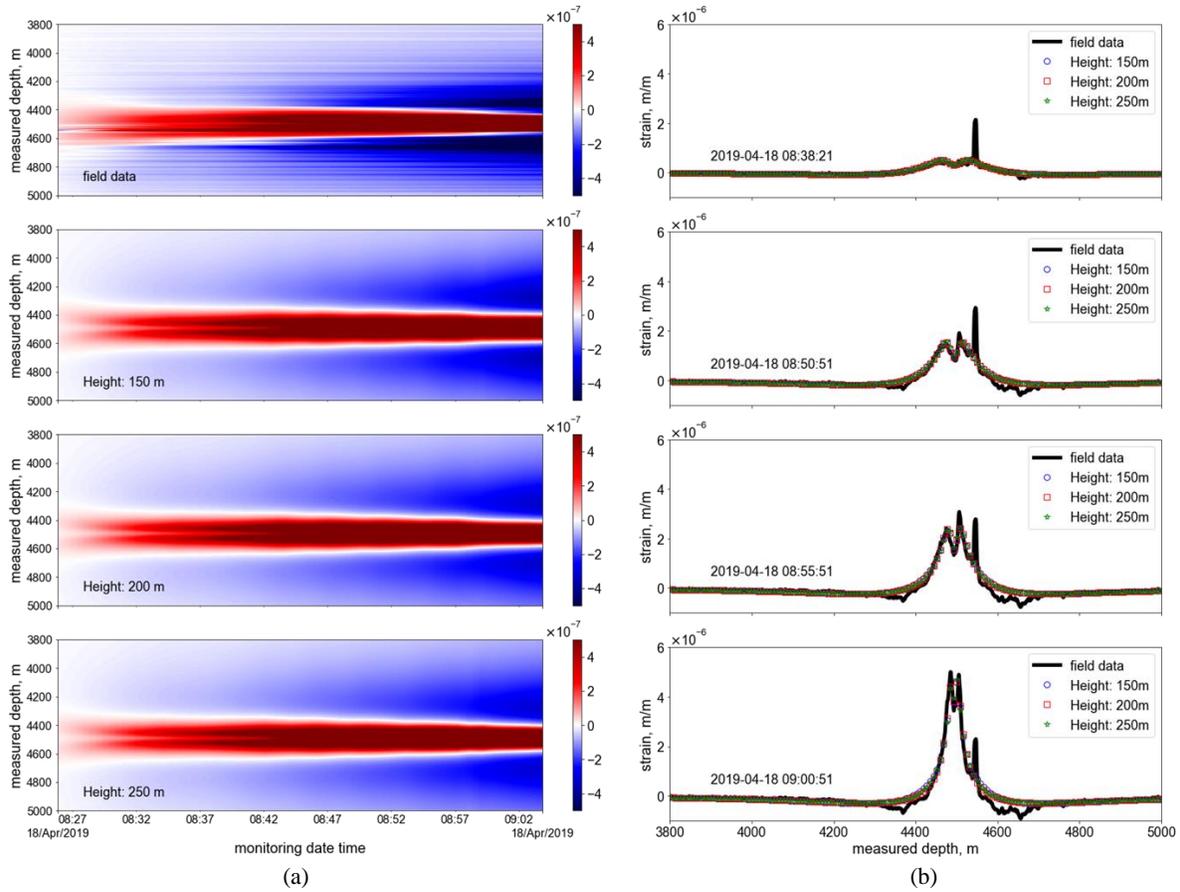

**Figure 11.** Comparison of field strain data and modeled strain data with different fracture heights. (a) strain waterfall plots; from top to bottom: field data, modeled data with 150 m fracture height, modeled data with 200 m fracture height, modeled data with 250 m fracture height. (b) strain profiles as a function of measured depth at different times; the specific time is labeled in each subplot. In each subplot, the black curve represents the field data, blue circle, red square, and green star represent the modeled data with 150 m fracture height, 200 m fracture height, and 250 m fracture height, respectively. The abnormal peak noise at about 4550 m is caused by the existing fracture in the previous stage.

In addition to fracture propagation velocity, we can also calculate other important parameters, such as fracture cross-section area and fracture volume, based on the inversion results. These parameters can help provide valuable information for evaluating the efficiency of hydraulic fracturing. As shown in **Figure 12**(a), the average fracture area does not differ significantly among the three cases with different fracture heights. Although higher fractures have a lower average width, their product (average width × height), i.e., the area, remains almost the same. On the other hand, **Figure 12**(b) displays the evolution of one-sided fracture volume and the volume of injected slurry. By comparing these two quantities, we can assess the efficiency of fracturing fluid in creating an effective fracture volume. We only calculate the fracture volume on one side where the monitoring well is located, because the fracture may not propagate symmetrically. There is also evidence from microseismic monitoring that the fracture is not symmetric to the perforation point in this case (Ugueto et al., 2021).

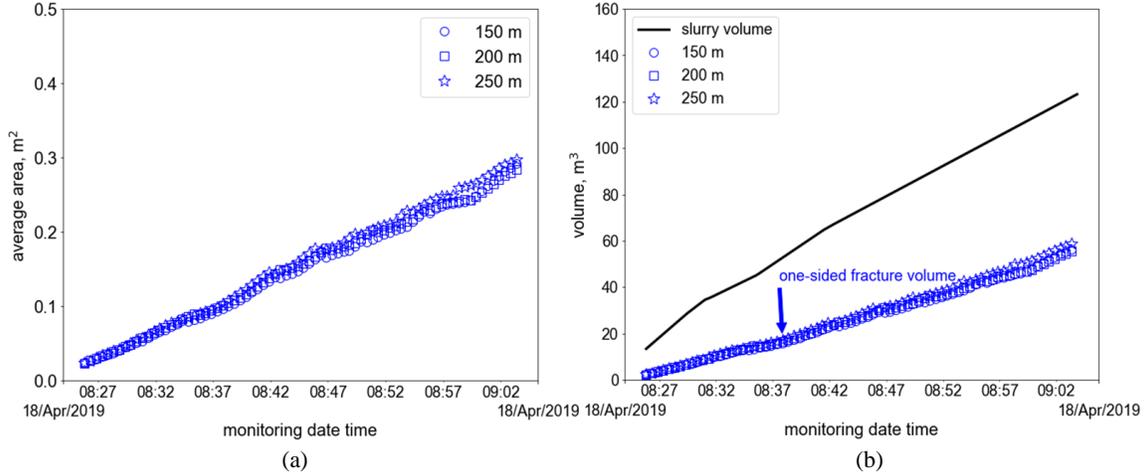

**Figure 12.** Average fracture cross-section area (a) and one-sided fracture volume as a function of monitoring time before fracture hit. The average cross-section area is calculated by $\bar{w} \times H$, where $\bar{w}$ is the averaged fracture width. The one-sided fracture volume is calculated by $\bar{w} \times H \times L/2$, where $L/2$ is the fracture half-length.

## 5 Conclusions and Recommendations

Low-frequency distributed acoustic sensing presents a groundbreaking opportunity to gain unprecedented insights into the mechanics of fluid-driven fracture propagation in porous media. Previous studies have either focused on qualitative interpretation through forward modeling or quantitative interpretation of the LF-DAS data after fracture hit through inverse modeling. In this study, we introduce a novel inversion approach that analyzes LF-DAS data before fracture hit, providing direct information on fracture tip advancement behavior and propagation velocity. Our method opens new avenues for understanding the intricate mechanics of hydraulic fracturing using the unique and precise LF-DAS data. The following key points can be obtained from this study:

1. A gradient-based inversion algorithm is developed to estimate the fracture tip advancement behavior using LF-DAS strain data before the fracture hit. The presented algorithm also allows for calculation of the average cross-section fracture area, fracture volume, and fracturing efficiency.
2. The performance of the algorithm is tested using a synthetic case with known reference solutions, which demonstrates that the estimated temporal evolutions of fracture half-length and propagation velocity match well with the reference solutions.
3. The inversion results, particularly the fracture half-lengths and propagation velocities, are not sensitive to the pre-defined fracture height. To estimate the fracture cross-section area and volume, an appropriate assumption of fracture width distribution may be required.
4. In this specific field case, the fracture propagates continuously in the early period, followed by an obvious intermittent advancement phenomenon. Such direct field observations help to understand the fracture propagation patterns and improve hydraulic fracturing models.
5. The results from the field case show a nearly linear increase in both fracture area and volume. However, further investigation and more field data are needed to understand how specific influencing factors such as fluid properties, formation properties, or operational parameters impact these observations.

To gain a deeper understanding of fluid-driven fracture propagation in porous media, we recommend more scientific field pilots with simple completion designs but various operational parameters to gather direct field observations. These observations can provide valuable insights for future hydraulic fracturing modeling and operations.

## Acknowledgements

The data used in this paper is downloaded from the NETL's Energy Data eXchange repository (https://edx.netl.doe.gov/group/gti-hfts-2).

## References

Adachi, J., Siebrits, E. M., Peirce, A., Desroches, J. 2007. Computer simulation of hydraulic fractures. *International Journal of Rock Mechanics and Mining Sciences*, *44*(5), 739-757.
Branch, M. A., Coleman, T. F., Li, Y. 1999. A subspace, interior, and conjugate gradient method for large-scale bound-constrained minimization problems. *SIAM Journal on Scientific Computing*, *21*(1), 1-23.
Bunger, A. P., Detournay, E. 2008. Experimental validation of the tip asymptotics for a fluid-driven crack. *Journal of the Mechanics and Physics of Solids*, *56*(11), 3101-3115.
Cao, T. D., Hussain, F., Schrefler, B. A. 2018. Porous media fracturing dynamics: stepwise crack advancement and fluid pressure oscillations. *Journal of the Mechanics and Physics of Solids*, *111*, 113-133.
Chen, B., Barboza, B. R., Sun, Y., Bai, J., Thomas, H. R., Dutko, M., Cottrell, M. et al. 2021. A review of hydraulic fracturing simulation. *Archives of Computational Methods in Engineering*, 1-58.
Chen, Y., Nagaya, Y., Ishida, T. 2015. Observations of fractures induced by hydraulic fracturing in anisotropic granite. *Rock Mechanics and Rock Engineering*, *48*, 1455-1461.
Cochard, T., Svetlizky, I., Albertini, G., Viesca, R., Rubinstein, S., Spaepen, F., Yuan, C., Denolle, M., Song, Y.Q., Xiao, L., Weitz, D., 2023. Unexpected Dynamics in the Propagation of Fracture Fronts.
Crouch, S. 1976. Solution of plane elasticity problems by the displacement discontinuity method. I. Infinite body solution. *International journal for numerical methods in engineering*, *10*(2), 301-343.
Garagash, D., Detournay, E. 2000. The tip region of a fluid-driven fracture in an elastic medium. *J. Appl. Mech.*, *67*(1), 183-192.
Geertsma, J., De Klerk, F. 1969. A rapid method of predicting width and extent of hydraulically induced fractures. *Journal of petroleum technology*, *21*(12), 1571-1581.
Ichikawa, M., Kurosawa, I., Uchida, S., Kato, A., Ito, Y., Takagi, S. de Groot, M. et al. 2019. Case study of hydraulic fracture monitoring using low-frequency components of DAS data. In *SEG Technical Program Expanded Abstracts 2019* (pp. 948-952). Society of Exploration Geophysicists.
Jin, G., Roy, B. 2017. Hydraulic-fracture geometry characterization using low-frequency DAS signal. *The Leading Edge*, *36*(12), 975-980.
Lecampion, B., Bunger, A., Zhang, X. 2018. Numerical methods for hydraulic fracture propagation: A review of recent trends. *Journal of natural gas science and engineering*, *49*, 66-83.
Lecampion, B., Desroches, J., Jeffrey, R. G., Bunger, A. P. 2017. Experiments versus theory for the initiation and propagation of radial hydraulic fractures in low-permeability materials. *Journal of Geophysical Research: Solid Earth*, *122*(2), 1239-1263.
Lhomme, T. P., De Pater, C. J., Helfferich, P. H. 2002. Experimental study of hydraulic fracture initiation in Colton sandstone. In *SPE/ISRM Rock Mechanics Conference*. OnePetro.
Li, X., Zhang, J., Grubert, M., Laing, C., Chavarria, A., Cole, S., Oukaci, Y. 2020. Distributed acoustic and temperature sensing applications for hydraulic fracture diagnostics. In *SPE Hydraulic Fracturing Technology Conference and Exhibition*. OnePetro.
Liu, D., Lecampion, B. 2022. Laboratory investigation of hydraulic fracture growth in Zimbabwe gabbro. *Journal of Geophysical Research: Solid Earth*, *127*(11), e2022JB025678.
Liu, Y., 2021. *Hydraulic fracture geometry characterization using low-frequency distributed acoustic sensing data: forward modeling, inverse modeling, and field applications.* Doctoral dissertation, Texas A&M University, College Station, Texas, USA.